\documentclass[submission]{SciPost}
\pdfoutput=1

\usepackage{neel}
\usepackage{cancel,comment}
\usepackage{shuffle}

\setcitestyle{numbers}
\begin{document}
	\begin{samepage}
		\begin{flushleft}\huge{Perturbative Unitarity Calls for An Action}\end{flushleft}
		\vspace{10pt}
		{\color{myviolet}\hrule height 1mm}
		
		\vspace*{10pt}
		\begin{flushleft}
		\large 	\textbf{Subhroneel Chakrabarti} and \textbf{Renann Lipinski Jusinskas}
		\end{flushleft}

		\begin{flushleft}
			\emph{\large Institute of Physics of the Czech Academy of Sciences \& CEICO\\ Na Slovance 2, Prague 182 21, Czech Republic.}
			
            \href{mailto:subhroneelc@gmail.com}{subhroneelc@gmail.com, } \href{mailto:renannlj@fzu.cz}{renannlj@fzu.cz}\\
		\end{flushleft}
		
		\begin{flushright}
		\end{flushright}

		\section*{Abstract}
		{In this work, we investigate the consistency of a perturbative definition of the S-matrix in a particular class of non-Lagrangian theories. We focus on the $p$-form theories proposed in \cite{Broccoli:2021pvv}, which are fully defined by “third-way” consistent equations of motion. Using the perturbiner method, we show that the unitarity is absent even at the tree level. We then pin down a unique modification of the equations of motion that restores unitarity. The trade-off is the reinstatement of an underlying Lagrangian, which we recognize as the higher-dimensional generalization of the Freedman-Townsend (FT) model. Finally, we discuss conserved currents in third-way theories and show they all follow from parent currents in the FT model. In particular, we point out the existence of a higher-ranked global symmetry, which signals that the FT model is compatible with the existence of brane-like charged objects in higher dimensions.}
		
	\end{samepage}
	
	\noindent\rule{\textwidth}{1pt}
	\pagecolor{white}
	\tableofcontents\thispagestyle{fancy}
	\noindent\rule{\textwidth}{1pt}

\section{Overview}
    There is multiple evidence for the existence of quantum field theories (QFTs) whose dynamics are not derivable from an action principle. Prominent examples are the Argyres-Douglas type theories \cite{Argyres:1995jj,Minahan:1996fg}, the six-dimensional $(2,0)$ superconformal field theories \cite{Witten:1995zh,witten:2008,Witten:2009at}, and the four-dimensional $\mathcal{N}=3$ theories \cite{Garcia-Etxebarria:2015wns,Aharony:2015oyb}. However, all of them are typically strongly coupled, argued to be inherently quantum \cite{Xie:2012hs,Wang:2015mra,Garcia-Etxebarria:2015wns} and non-perturbative. They do not have a classical limit, a feature compatible with the lacking action. As superconformal field theories, nevertheless, they are amenable to bootstrap techniques that enable the computation of certain observables \cite{Beem:2015aoa,Beem:2014zpa,Lemos:2016xke}. Alternatively, given their natural embedding into string theory, one can even attempt a top-down construction starting from a suitable string theoretic set-up \cite{Heckman:2018jxk,Aharony:2016kai}. In fact, a large class of Argyres-Douglas type theories can also be obtained by compactifying $(2,0)$-theories \cite{Xie:2012hs,Wang:2015mra}.  It is then natural to ask whether one can define weakly coupled QFTs that cannot be derived from an action principle. A related intriguing question was investigated in the case of $\mathcal{N}=3$ theories in $d=4$ flat spacetime. They were previously argued to be nonexistent \cite{Weinberg:2000cr}, but the argument implicitly assumed both the existence of an action and a suitable weak-coupling limit where perturbation theory makes sense. Their eventual construction has bypassed this no-go result by virtue of being intrinsically strongly coupled, and therefore non-Lagrangian.

    Interestingly, modern methods for scattering amplitude computation tend to bypass the need for an action functional. These techniques are based on the fact that tree-level amplitudes can be recursively derived on the mass-shell \cite{Travaglini:2022uwo,Brandhuber:2022qbk,Bern:2022wqg}, with loop computations following from unitary cuts \cite{Bern:1994cg,Bern:1994zx,Bern:2011qt}. However, there are few examples of perturbative S-matrix computations in theories that do not have an action, even though such an input is ultimately not necessary. We know, for instance, of certain field theories with equations of motion that cannot be derived from an action principle. A typical example is the Navier-Stokes equation, which was diagrammatically described a long time ago \cite{Wyld:1961gqg}. It has been generalized to a non-Abelian theory in \cite{Cheung:2020djz} (see also \cite{Escudero:2022zdz}), including tree-level computations and even a double-copy analysis. However, since the Navier-Stokes equation describes dissipative phenomena, the isolated system is not unitary. In addition, the non-Abelian interaction introduced in \cite{Cheung:2020djz} leads to scattering trees with asymmetric external states, and even the usual cyclicity of a color-ordered amplitude is absent.

    Recently, a novel set of equations of motion was proposed for a set of colored self-interacting $(d-2)$-form potentials in $d$ spacetime dimensions \cite{Broccoli:2021pvv}. None of them can be derived from an action principle. Nonetheless, they are shown to be consistent in what is dubbed the ``third-way'' of constructing gauge theories \cite{Bergshoeff:2014pca,Bergshoeff:2015zga,Bergshoeff:2018luo,Arvanitakis:2015oga,Afshar:2019npk,Ozkan:2018cxj}. It may seem that a classical analysis based on the equations of motion is the best that one can do with these interacting field theories. However, it has been long known that classical equations of motion are enough to extract the knowledge of tree-level scattering amplitudes. One such algorithm is the perturbiner method \cite{Rosly:1996vr,Rosly:1997ap}, which was later rediscovered in \cite{Mafra:2015gia,Mafra:2015vca,Lee:2015upy}. Based on multi-particle solutions to equations of motion, the method has been applied to several different theories, even being modified to compute $1$-loop amplitudes and correlators \cite{Lee:2022aiu,Gomez:2022dzk,Gomez:2024xec}. Therefore, it is certainly possible to compute a putative S-matrix for these theories at tree-level and investigate if they satisfy the required properties of a perturbative QFT.
    
    In this work, we use the perturbiner method to compute and analyze tree-level scattering amplitudes of third-way interacting $p$-form theories of \cite{Broccoli:2021pvv}. Our analysis reveals two crucial facts: 
    \begin{enumerate}
      \item All third-way equations of motion introduced in \cite{Broccoli:2021pvv} break tree-level unitarity.
      \item Fixing tree-level factorization uniquely leads to a higher-dimensional realization of the Freedman-Townsend model (FT) \cite{Freedman:1980us}.
    \end{enumerate}

    \noindent As far as  these theories are concerned, our results demonstrate that one is forced to go back to Lagrangian dynamics when demanding unitarity. It would be tempting to conjecture that weakly interacting, well-defined QFTs must allow a classical description in terms of an action, but a more rigorous analysis is necessary.

    The rest of this work is outlined as follows. In section \ref{sec:EOM} we review the third-way equations of motion and its gauge fixing while establishing our notation and conventions. In section \ref{sec:perturbiner} we set up the perturbiner method for computing tree-level S-matrix starting from the gauge-fixed equation of motion. Section \ref{sec:scattering} contains the key result of this article, where we demonstrate that the third-way equations all fail to give us a unitary S-matrix. We propose a correction that restores unitarity and show that the modified equations of motion are now the Euler-Lagrange equations of the FT model. We also point out that all extensions of third-way equations involving a lower-ranked potential also fail to be unitary, albeit for a simpler reason. In addition to investigating the tree-level S-matrix, in section \ref{sec:symmetry} we discuss global symmetries and conserved currents in third-way theories. Specifically, we show that the third-way theories possess a conserved higher-form global charge. This is suggestive of extended, brane-like objects charged under this symmetry. All such conserved charges follow from the underlying gauge symmetry of the FT model. In section \ref{sec:final} we conclude with brief comments on several open questions and directions that naturally arise from this work.

\section{The equations of motion} \label{sec:EOM}

    Consider a non-Abelian flat connection $G=0$, with
    \begin{equation}\label{eq:EOM}
	G= \dd \tilde{H} + \kappa  \tilde{H} \wedge \tilde{H}.
	\end{equation}
    Here $\kappa$ is the coupling constant and $\tilde{H} = \tilde{H}^I T_I$. The group generators $T_I$ satisfy $[T_I,T_J]= f^{K}_{IJ} T_K$, with structure constant $f_{JK}^{I}$. Lie-valued wedge products implicitly involve a commutator of the group generators. Notice that by taking the exterior derivative on both sides of \eqref{eq:EOM}, it is easy to show that
    \begin{eqnarray}
    \mathrm{d} G^I &=& \mathrm{d} (\mathrm{d} \tilde{H}^I + \kappa f^{I}_{JK} \tilde{H}^J \wedge \tilde{H}^K), \nonumber \\
    &=& 2 \kappa f^{I}_{JK} \mathrm{d}\tilde{H}^J \wedge \tilde{H}^K, \nonumber \\
    &=& 2 \kappa f^{I}_{JK} G^J \wedge \tilde{H}^K.
    \end{eqnarray}
    We have used the Jacobi identity  $f^{I}_{JK} f^{J}_{LM}+f^{I}_{JL} f^{J}_{MK}+f^{I}_{JM} f^{J}_{KL} =0$ to go from the second to the third line.

    Equation \eqref{eq:EOM} becomes non-trivial when $\tilde{H}$ is seen as the dual field-strength of a $(d-2)$-form $C$ in a $d$-dimensional Minkowski spacetime (with $d \geq 4$), i.e., $\tilde{H}= \star H$ and $H=\mathrm{d} C$. This was the case analyzed in \cite{Broccoli:2021pvv}, and it is possible to show that such an equation cannot follow from an action principle. Since $\mathrm{d} G$ vanishes on the mass shell ($\mathrm{d} G \propto G$), the equation of motion $G=0$ is said to be third-way consistent. From a field theory perspective, however, we noticed that scattering amplitudes derived using this model lack basic physical properties, such as factorization. We will come back to this in section \ref{sec:scattering}.
    
    In order to remedy this, we have analyzed possible variations of the field-strength definition and their impact on tree-level amplitudes. It turns out there is a unique addition to $H$ that preserves the third-way character of the theory. Instead of $H=\mathrm{d} C$, consider
    \begin{equation}
        H=\mathrm{d} C + 2 \alpha \kappa \tilde{H} \wedge C, \label{eq:nonAb-FS}
    \end{equation}
    where $\alpha$ is a numerical parameter. For $\alpha \neq 0$, the intrinsic definition of $H$ in \eqref{eq:nonAb-FS} leads to an infinite expansion in $\kappa$, though with an interesting property. Since $\tilde{H}$ appears to play the role of a connection, consider also the non-Abelian gauge transformation
    \begin{equation}
        \delta C = \dd \Lambda +2 \alpha \kappa \tilde{H} \wedge \Lambda,
    \end{equation}
   where the gauge parameter $\Lambda$ is a Lie algebra-valued $(d-3)$-form. It is easy to show that
   \begin{equation}
       \delta H = \kappa  ( G \wedge \Lambda + \delta \tilde{H} \wedge C ),
   \end{equation}
   when $\alpha=1$. This implies that the field-strength $H$ is gauge invariant on-shell.
  
    Our next goal is to investigate classical multi-particle solutions of $G=0$, with $H$ defined in \eqref{eq:nonAb-FS}. It will often be more convenient to explicitly work with components. We denote the flat metric by $\eta_{\mu \nu}$, which is used to raise and lower vector indices, and the Levi-Civita symbol by $\epsilon_{\mu_1 \cdots \mu_d}$. We then introduce the notation $\{\mu\}=\mu_1 \cdots \mu_{d-2}$ for a string of antisymmetrized $(d-2)$ indices, such that 
    \begin{equation}
    H_{\nu \{\mu\}} = \frac{1}{(d-1)} \left(\partial_\nu C_{\{\mu\}} + 2 \alpha \kappa [\tilde{H}_\nu , C_{\{\mu\}}] \right) - (\nu \leftrightarrow \mu_i).
    \end{equation}
    In terms of the dual field-strength, we have $H_{\nu \{\mu\}}= \tilde{H}^{\rho} \epsilon_{\rho \nu \{\mu\}}$, and
    \begin{equation}
        \tilde{H}^{\rho} = \frac{1}{(d-1)!}\epsilon^{\rho \nu \{\mu\}} H_{\nu \{\mu\}},
    \end{equation}

    Equation \eqref{eq:EOM} can be simply cast as
    \begin{equation} \label{eq:EOM-Htilde}
        \partial^{\mu} \tilde{H}^{\nu} -\partial^{\nu} \tilde{H}^{\mu} + 2 \kappa [\tilde{H}^{\mu},\tilde{H}^{\nu}] = 0,
    \end{equation}
    We then fix the form symmetry of $C$ through the transversal gauge $\dd^\dagger C = 0$, which in components translates to $\partial^{\mu_1} C_{\mu_1\cdots \mu_{d-2}} = 0$, and leads to the equation of motion
	\begin{equation} \label{eq:box-EOM}
		\Box C_{\bl{\mu}}  = \kappa (d-1) [ H_{\nu \{\mu\}},\tilde{H}^{\nu}] - 2 \alpha \kappa \partial^\nu [\tilde{H},C]_{\nu\{\mu\}}.
	\end{equation}
    Here, we have $\Box=\eta^{\mu \nu} \partial_\mu \partial_\nu$, and
    \begin{equation}
     [\tilde{H},C]_{\nu\{\mu\}} = [\tilde{H}_\nu, C_{\{\mu\}}]  - (\nu \leftrightarrow \mu_i).
    \end{equation}
    Next, we discuss the perturbiner solutions of \eqref{eq:box-EOM}.
    
\section{Classical multi-particle solutions} \label{sec:perturbiner}

    In the free case ($\kappa=0$), the solutions of \eqref{eq:box-EOM} correspond to single-particle states of the form
    \begin{equation}
        C_{\{\mu\}} =  c_{p\{\mu\}}e^{\mathrm{i}k_p\cdot x} T^{I_p}, \label{eq:single}
    \end{equation}
    with momentum $k_{p}$ satisfying the massless condition $k_p^2=0$. We denote by $c_p$  the color-stripped polarization of the particle labeled by the \textit{letter} $p$, and corresponding field-strength $h_{p}$. The transversal gauge implies that $k_{p}^{\mu_1} c_{p \mu_1 \ldots \mu_{d-2}}=0$. The generator $T^{I_p}$ takes care of the group structure of the solution.

    The single-particle solutions \eqref{eq:single} are the building blocks of a formal multi-particle solution of the full equation of motion \eqref{eq:box-EOM}, dubbed perturbiner. In order to derive it, we start with the ansatz
    \begin{equation}
        C_{\{\mu\}} = \sum_{P}\mathcal{C}_{P \{\mu\}}e^{\mathrm{i}k_{P}\cdot x}T^{I_{P}}. \label{eq:ansatz}
    \end{equation}
    The \textit{word} $P=p_1 p_2 \ldots p_n$ is a sequence of single-particle labels, such that $k_{P}=k_{p_1}+\ldots+k_{p_{n}}$ and $T_{I_{P}}=T_{I_{1}}\cdots T_{I_{n}}$. The sum goes over non-empty words $P$.  One-letter currents are simply given by their single-particle counterparts. For convenience, we introduce
    \begin{eqnarray}
        H_{\nu \{\mu\}} &=& \mathrm{i} \sum_{P}\mathcal{H}_{P \nu \{\mu\}}e^{\mathrm{i}k_{P}\cdot x}T^{I_{P}},\\
        \mathcal{H}_{P \nu \{\mu\}} &=& \frac{1}{(d-1)} \left( k_{P \nu} \mathcal{C}_{P \{\mu\}} +2 \alpha \kappa \sum_{P=QR} \left[ \tilde{ \mathcal{H}}_{Q\nu} \mathcal{C}_{R\{\mu\}}   -(Q\leftrightarrow R)\right] \right)- (\nu \leftrightarrow \mu_i),
    \end{eqnarray}
    and the sum in $\mathcal{H}_{P \nu \{\mu\}}$ goes over all non-empty sub-words $Q$ and $R$ such that $P=QR$. The multi-particle currents $\mathcal{C}_{P \{\mu\}}$ are recursively determined when plugging the ansatz back in the equation of motion, and it is straightforward to show that
	\begin{equation} \label{eq:C-recursion}
	    \mathcal{C}_{P \{\mu\}}= -\frac{\kappa}{s_{P}}\sum_{P=QR} \left[(d-1)\tilde{\mathcal{H}}^{\nu}_{Q} \mathcal{H}_{R \nu \{\mu\}} + 2 \alpha k_{P}^{\nu} (\tilde{\mathcal{H}}_{Q} \mathcal{C}_{R})_{\nu\{\mu\}} -(Q\leftrightarrow R) \right]. 
	\end{equation}
	with $s_{P}=k_{P}^{2}$, and $\tilde{\mathcal{H}}^{\nu}_{P}$ denoting the dual of $\mathcal{H}_{P \nu \{\mu\}}$. It also follows from the equation of motion that the multi-particle currents $\mathcal{C}_{P}$ satisfy the shuffle identities,
    \begin{equation}
        \mathcal{C}_{Q\shuffle R}^{\mu}=0, \label{eq:shuffle-id}
    \end{equation}
    with the operation $Q\shuffle R$ denoting the sum over all possible shuffles between the words $Q$ and $R$. Equation \eqref{eq:shuffle-id} implies that $C_{\{\mu\}}$ in \eqref{eq:ansatz} is Lie-algebra valued, constituting a formal solution of the non-linear equation \eqref{eq:box-EOM} when the multi-particle currents are given by \eqref{eq:C-recursion}.

 \section{Tree level unitarity of third-way theories}\label{sec:scattering}

    The physical interpretation of the multi-particle currents $\mathcal{C}_{P \{\mu\}}$ is well known, corresponding to a tree-level amplitude with one off-shell leg. The transversal gauge on $C_{\{\mu\}}$ is then manifested as a Ward identity,
    \begin{equation}
        k^{\mu_1}_P \mathcal{C}_{P \mu_1\cdots \mu_{d-2}} = 0. \label{eq:gauge-multi}
    \end{equation}
    Indeed, when attaching to $\mathcal{C}_{P \{\mu\}}$ an external single-particle polarization, it is possible to define the tree-level color-ordered (partial) amplitude of the model by imposing momentum conservation:
    \begin{equation}
      A(1,\ldots,n)= \lim_{s_{2...n}\to0}c_{1}^{\{\mu\}}s_{2...n} \mathcal{C}_{2\ldots n \{\mu\}}.\label{eq:amp-tree}  
    \end{equation}
    The gauge invariance of $A(1,\ldots,n)$ under residual single-particle transformations can be checked using \eqref{eq:gauge-multi}. Furthermore, the partial amplitudes \eqref{eq:amp-tree} satisfy Kleiss-Kuijf relations \cite{Kleiss:1988ne}, which follow from the shuffle identities \eqref{eq:shuffle-id}.

    \subsection{Lower-point examples and factorization}
    
    The first non-trivial example is the three-point case, $A(1, 2, 3)$, which requires the computation of the two-particle current:
    \begin{equation} 
	    \mathcal{C}_{23 \{\mu\}}= -\frac{\kappa}{s_{23}}\left[(d-1)\tilde{h}^{\nu}_{2} h_{3 \nu \{\mu\}} + 2 \alpha k_{23}^{\nu} (\tilde{h}_2 c_3)_{\nu\{\mu\}}\right]  -(2\leftrightarrow 3) .
	\end{equation}
    It is then easy to show that
    \begin{equation}
        A(1,2,3)= 2 \kappa (d-1)  \left(\tilde{h}^{\nu}_{3} h_{2 \nu \{\mu\}}  c_{1}^{\{\mu\}}  + \alpha \tilde{h}_2^{\nu }h_{1 \nu \{\mu\}} c_3^{\{\mu\}} + \alpha \tilde{h}_{1}^{\nu} {h}_{3\{\mu\}} c_2^{\{\mu\}}\right).
    \end{equation}
    For $\alpha=1$, the partial amplitude defined as in \eqref{eq:amp-tree} is cyclic,
    \begin{equation}
        A(1,2,3)=A(2,3,1)=A(3,1,2).
    \end{equation}
    In flat space, Lagrangian field theories are guaranteed to have cyclic interaction vertices using momentum conservation. This is not the case, for instance, in $AdS$, where the cyclicity of the gluon amplitude is broken up to a boundary contribution \cite{Armstrong:2022mfr}. The $\alpha=0$ case considered in \cite{Broccoli:2021pvv} does not immediately satisfy cyclicity, and the definition \eqref{eq:amp-tree} would need to be supplemented. 

    The four-point ordered amplitude is written in terms of the three-particle current,
	\begin{multline}
	    \mathcal{C}_{234 \{\mu\}}= \frac{2 \kappa}{s_{234}}\Big\{
        (d-1)( \tilde{\mathcal{H}}^{\nu}_{34} h_{2 \nu \{\mu\}}  +\tilde{h}^{\nu}_{4} \mathcal{H}_{23 \nu \{\mu\}})
        \\  + \alpha k_{234}^{\nu} [(\tilde{\mathcal{H}}_{34} c_{2})_{\nu\{\mu\}} - (\tilde{h}_2 \mathcal{C}_{34})_{\nu\{\mu\}}  +  (\tilde{h}_4 \mathcal{C}_{23})_{\nu\{\mu\}} -  (\tilde{\mathcal{H}}_{23} c_{4})_{\nu\{\mu\}}]\Big\}. \label{eq:B-recursion_4pt}
	\end{multline} 
	which then leads to
    \begin{multline}
	    A(1,2,3,4)= 2 \kappa (d-1) \left(\tilde{\mathcal{H}}^{\nu}_{34} h_{2 \nu \{\mu\}} c_1^{\{\mu\}}  +\alpha \tilde{h}_{2}^{\nu} h_{1\nu\{\mu\}}  \mathcal{C}_{34}^{\{\mu\}} + \alpha \tilde{h}_{1}^{\nu} \mathcal{H}_{34 \nu\{\mu\}}  c_{2}^{ \{\mu\}}\right) \\ - 2 \kappa (d-1) \left(\tilde{\mathcal{H}}^{\nu}_{23} h_{4 \nu \{\mu\}} c_1^{\{\mu\}}  +\alpha \tilde{h}_{4}^{\nu} h_{1\nu\{\mu\}}  \mathcal{C}_{23}^{\{\mu\}} + \alpha \tilde{h}_{1}^{\nu} \mathcal{H}_{23 \nu\{\mu\}}  c_{4}^{ \{\mu\}}\right).
	\end{multline} 

    In order to analyze the factorization of the four-point amplitude, it is more enlightening to write it explicitly in terms of $\mathcal{C}_{23}$ and $\mathcal{C}_{34}$. After some algebra, we obtain
    \begin{multline}
	    A(1,2,3,4)= s_{34} \mathcal{C}_{34}^{\{\mu\}}\mathcal{C}_{12 \{\mu\}} + s_{23} \mathcal{C}_{23}^{\{\mu\}}\mathcal{C}_{41 \{\mu\}} \\ +2(1-\alpha) \kappa \mathcal{C}_{34}^{\{\mu\}} \left[(d-1)\tilde{h}^{\nu}_{1} h_{2 \nu \{\mu\}} + k_{12}^{\nu} (\tilde{h}_2 c_1)_{\nu\{\mu\}}\right] \\
        -2(1-\alpha) \kappa \mathcal{C}_{23}^{\{\mu\}} \left[(d-1)\tilde{h}^{\nu}_{1} h_{4 \nu \{\mu\}} +k_{14}^{\nu} (\tilde{h}_4 c_1)_{\nu\{\mu\}}\right]. \label{eq:4pt-amp}
	\end{multline}
    For an arbitrary value of $\alpha$, this amplitude is not even cyclic. Only when $\alpha = 1$, we see that the four-point amplitude is manifestly cyclic and correctly factorizes in the $s_{23}$ and $s_{34}$ channels. For example,
    \begin{equation}
        \left.\vphantom{\frac{1}{1}} A(1,2,3,4)\right|_{s_{34}\to 0} = \frac{1}{s_{34}} A(1,2,*) A (*,3,4),
    \end{equation}
    with $*$ denoting the state propagating in the channel.	

    \subsection{Freedman-Townsend Model}

    The proposed ``correction'' to  the field-strength \eqref{eq:nonAb-FS} restores the tree-level unitarity of the theory when $\alpha=1$. As it turns out, this ends up breaking the  \textit{non-Lagrangian} character of the equation of motion $G=0$, which can then be derived from an action principle. Indeed, the system is identified with a higher-dimensional realization of the FT model, and we will briefly summarize its key details in higher dimensions. 
    
    In its first-order form, the FT action is given by
    \begin{equation}
    S = \int \left( \frac{1}{2} A^I \wedge \star A^I - A^I \wedge \mathrm{d} C^I - \kappa f_{IJK} A^I \wedge A^J \wedge C^K  \right), \label{eq:FT-action}
    \end{equation}
    where $C= C^I T_I$ is a $(d-2)$-form and $A= A^I T_I$ is a $1$-form. 

    The Euler-Lagrange equations for this action are given by
    \begin{eqnarray}
    \star A &=& \mathrm{d} C +2 \kappa A \wedge C, \\
    \mathrm{d} A + \kappa A \wedge A & = & 0,
    \end{eqnarray}
    and it is straightforward to see that the local transformations with parameter $\Lambda$,
    \begin{eqnarray}
    \delta C &=&  \mathrm{d} \Lambda + 2 \kappa A \wedge \Lambda, \\
    \delta A & = & 0,
    \end{eqnarray}
    leave the action \eqref{eq:FT-action} invariant up to a boundary contribution.

    At this point we readily recognize the modified equation introduced in section \ref{sec:EOM} with $\alpha=1$, which restores unitarity to the third-way equations. If one integrates out the $1$-form $A$ from \eqref{eq:FT-action}, one obtains the second-order realization of the FT action. In spite of having an infinite number of terms, the associated equation of motion is simply a flat-connection condition for $\Tilde{H}$. Indeed, both the third-way and the FT model equations of motion are of this kind, with the crucial difference being the  definition of this flat connection in terms of the potential $C$.

    \subsection{Inclusion of other lower-ranked form fields}

    One interesting feature of the third-way models proposed in \cite{Broccoli:2021pvv} is the inclusion of a $(d-3)$-form potential  $c = c^i T_i$, with group generators not necessarily the same as in $C=C^I T_I$. Its equation of motion is given in terms of the dual field strength $\tilde{h}^i = \star \dd c^i$, with
    \begin{equation}
            \mathrm{d} \tilde{h}^i= \tilde{\kappa} \, \mathcal{F}^{i}_{JKL} \tilde{H}^J \wedge \tilde{H}^K \wedge \tilde{H}^L. \label{eq:lower-eom}
    \end{equation}
    
    Here, third-way consistency requires that the structure constants $\mathcal{F}^{i}_{JKL}$ satisfy 
    \begin{equation} \label{eq:new_Jacobi}
        \mathcal{F}^i_{J[MN} f^J_{KL]}=0.
    \end{equation}
    This can be explicitly checked by acting with the exterior derivative on both sides and making use of the Jacobi-like identity given in \eqref{eq:new_Jacobi}. In a similar manner, it is also possible to add $p$-forms, with $p<(d-2)$. For more details and examples, we refer the readers to \cite{Broccoli:2021pvv}. 
    
    Without further deforming the field-strength in \eqref{eq:nonAb-FS}, the $(d-3)$-form potential $c^i$ plays the role of a probe field, in the sense that its dynamics is affected by $H$ without any back-reaction. From the perspective of the S-matrix, any attempt to define tree-level scattering amplitudes would have nonphysical factorization properties. This suggests that, while this coupling to a lower-ranked form is a third-way consistent classical equation of motion, it does not lead to unitary quantum dynamics. However, the reason for non-unitarity for these classes of theories is more obvious - it is simply the fact that at the level of dynamics and interaction, the two potentials are not on an equal footing. One can easily see that this is the case for \textit{all} the third-way consistent theories involving lower-ranked forms introduced in \cite{Broccoli:2021pvv}.

\section{Symmetries and Conserved Currents} \label{sec:symmetry}
	
	The vanishing of equation \eqref{eq:EOM} implies a $(d-2)$-form conserved current \cite{Gaiotto:2014kfa}. Starting with
	\begin{equation}\label{eq:new_j}
		J = \kappa  \star (\tilde{H} \wedge   \tilde{H}),
	\end{equation}
	it is clear that $\dd \star J = 0$ using $G=0$. Indeed, the current is co-exact on-shell. In this case, the conserved charge always vanishes on-shell by Stokes' law in ordinary flat spacetime, which is being considered here. Nonetheless, this charge will be non-zero once the theory is put on a spacetime with non-contractible $(d-2)$-cycle. This symmetry is, therefore, reminiscent of winding symmetry of compact bosons. In the free theory, this current is identically zero on-shell since the current is proportional to the coupling constant $\kappa$. Note that similar higher-form currents can be constructed for all of the classical equations introduced in \cite{Broccoli:2021pvv} that includes an additional lower-ranked form. Since they all lead to non-unitary theories, we will not provide the expression for those currents here. Of course, there is also the usual current present in any $p$-form theory, $\mathcal{J} = \star (\mathrm{d} C)$, which is trivially conserved. \\
	
Given a higher-form theory, typically one can just look at the fluxes or the conserved charges to figure out the dimensionality of the charged objects, which we will call \textit{branes}.

	For the full interacting theory, we use the $(d-2)$-form conserved current of \eqref{eq:new_j}. In flat space, the corresponding charge $Q^I$ is always zero on-shell. Nevertheless, it can be used to formally infer the dimensionality of the objects that should be charged under this symmetry. In this case, we have
    \begin{equation} \label{eq:charge}
		Q^I = \int_{\Sigma_2} \star J^I.
	\end{equation}
	To find the dimension of the object with charge $Q^I$, we need to look at the Poincar\'e dual of the closed surface $\Sigma_2$. This will be a sub-manifold of dimension $(d-1) - 2 = d-3$, which is the same dimensionality of the electrically charged brane in the free theory. Similarly, the dimension of the magnetically charged object is a $(-1)$-brane. This magnetically charged object can be interpreted as an instanton, just like the $D(-1)$-brane in string theory \cite{Gibbons:1995vg,Gubser:1996wt}. But unlike in string theory, these branes are not dynamical objects.

    Recall that the equations of motion of the third-way theories and the FT model can be traced back to a flat connection, only differing in their field-strengths $H$. However, the conservation of $J$ is oblivious to the definition of $\Tilde{H}$, and the current \eqref{eq:new_j} is also conserved in the FT model. It is identified with the global part of the gauge symmetry, when $\Lambda$ is a constant. We are not aware of any prior discussions of this higher global symmetries in the FT model, and it would be interesting to understand them better.

    Finally, there is a candidate conserved stress-tensor for the third-way theory \cite{Broccoli:2021pvv}, given by
    \begin{equation}
        T_{\mu \nu}=\tilde{H}_\mu^I \tilde{H}_\nu^I-\frac{1}{2} g_{\mu \nu} \tilde{H}_\rho^I \tilde{H}^{I \rho}.
    \end{equation}
    Once again, the conservation of $T_{\mu \nu}$ follows from $G=0$, and the precise definition of the $\Tilde{H}$ is irrelevant. Indeed, this proposal for $T_{\mu \nu}$ was drawn from the minimally coupled FT action.
    
    More generally, the set of global symmetries of the third-way equations of motion have all an analogous version in the FT model, where the corresponding currents can be systematically derived using textbook field-theory methods. 

\section{Final remarks} \label{sec:final}

    The main third-way theory introduced in \cite{Broccoli:2021pvv} shares several features with the Freedman-Townsend model. However, the specific difference in the definition of $H$ in \eqref{eq:nonAb-FS} that renders the third-way equations  non-Lagrangian also forces the theory to lose unitarity, even at tree-level. Therefore, this class of third-way theories, while perfectly valid as classical equations of motion, cannot lead to a sensible QFT. Let us highlight that the perturbiner was simply the tool we decided to use here. The outcome of the analysis would be the same using different methods based on on-shell constructibility. 
    
    Since the interaction term in the third-way equation does not involve additional degrees of freedom, it is natural to question whether this loss of unitarity could suggest dissipative behavior. In fact, given their similarity with non-Abelian Chern-Simons equations for the field $\Tilde{H}$, it would be interesting to investigate whether they could be describing a non-Abelian, higher-dimensional generalization of the shallow-wave equations in fluid dynamics. This is motivated by recent results in the Abelian case \cite{Tong:2022gpg}.
    
    The third-way equations involving additional lower-ranked forms fail to be unitary for a clearly different reason. There is an obvious asymmetry in the dynamics of different degrees of freedom, with the underlying $(d-2)$-form potential unaware of lower-ranked forms. Furthermore, the dynamics of the lower-ranked forms also describe conserved currents with the same rank. While we did not investigate possible ways to amend this lack of unitarity, this could lead to an interesting extension of the FT model with extra, lower-ranked fields. It would also offer a fertile ground to explore the fate of the higher-form symmetry of the extended model.

    Relatedly, it is also worth understanding the higher-form global symmetry in higher-dimensional FT theory, which acts non-trivially on spacetime with non-contractible $(d-2)$-cycles. It is known that this particular symmetry cannot be spontaneously broken following the higher-form generalizations of the Coleman-Mermin-Wagner theorem \cite{Gaiotto:2014kfa,Lake:2018dqm}. So there are no interesting phase transitions associated with this symmetry. Nonetheless, it is compelling to better understand the role of the extended charged objects. This is especially interesting because one of them is instanton-like, which typically captures non-perturbative aspects of a given theory. Moreover, the equivalence of the Freedman-Townsend (FT) model to principal chiral sigma models in $d=4$ is well-established \cite{Buchbinder:2024tui}, and finding an interpretation of the charged extended objects in this context would be enticing.

    We would like to conclude on a speculative note. Suppose we have a physically consistent Lagrangian QFT at a fixed point with S-matrix $\mathcal{S}_0$. Now we \emph{define} a new S-matrix perturbatively around $\mathcal{S}_0$, with
    \begin{equation}
        \mathcal{S} = \mathcal{S}_0 + \kappa \, \mathcal{S}_1 + O(\kappa^2) \;.
    \end{equation}
    We conjecture that if $\mathcal{S}$ satisfies unitarity and other general expected properties of the S-matrix, then it must also admit an action-based derivation. Our results offer but pieces of evidence to support this. However, we  believe that a systematic investigation aiming to (dis)prove this proposition should deepen our understanding of the space of quantum field theories. Given the fate of third-way interacting $p$-form theories, it would be fascinating to ascertain the (non)existence of weakly coupled self-interacting $p$-form fields that are consistent as a quantum theory, in particular because of their proliferation in string theory.

\section*{Acknowledgment}

We would like to thank Madhusudhan Raman for comments and suggestions. SC would like to thank TIFR (Mumbai), IMSc (Chennai), SNBNCBS (Kolkata), and IACS (Kolkata) for hospitality and an opportunity to present initial results of this work. SC acknowledges partial financial support from the European Structural and Investment Funds and the Czech Ministry of Education, Youth, and Sports (project FORTE CZ.02.01.01/00/22\_008/0004632) and the Grant Agency of Czech Republic under the grant EXPRO 20-25775X. RLJ acknowledges the financial support from the Czech Academy of Sciences under the project number LQ100102101.

\bibliographystyle{JHEP}
\bibliography{refs.bib} 

@article{Broccoli:2021pvv,
    author = "Broccoli, Matteo and Deger, Nihat Sadik and Theisen, Stefan",
    title = "{Third Way to Interacting p-Form Theories}",
    eprint = "2103.13243",
    archivePrefix = "arXiv",
    primaryClass = "hep-th",
    doi = "10.1103/PhysRevLett.127.091603",
    journal = "Phys. Rev. Lett.",
    volume = "127",
    number = "9",
    pages = "091603",
    year = "2021"
}

@article{Bergshoeff:2015zga,
    author = "Bergshoeff, Eric and Merbis, Wout and Routh, Alasdair J. and Townsend, Paul K.",
    title = "{The Third Way to 3D Gravity}",
    eprint = "1506.05949",
    archivePrefix = "arXiv",
    primaryClass = "gr-qc",
    doi = "10.1142/S0218271815440150",
    journal = "Int. J. Mod. Phys. D",
    volume = "24",
    number = "12",
    pages = "1544015",
    year = "2015"
}

@article{Bergshoeff:2014pca,
    author = "Bergshoeff, Eric and Hohm, Olaf and Merbis, Wout and Routh, Alasdair J. and Townsend, Paul K.",
    title = "{Minimal Massive 3D Gravity}",
    eprint = "1404.2867",
    archivePrefix = "arXiv",
    primaryClass = "hep-th",
    reportNumber = "RUG-CTN-2014-67, DAMTP-2014-20, MIT-CTP-4544",
    doi = "10.1088/0264-9381/31/14/145008",
    journal = "Class. Quant. Grav.",
    volume = "31",
    pages = "145008",
    year = "2014"
}

@article{Bergshoeff:2018luo,
    author = "Bergshoeff, Eric A. and Merbis, Wout and Townsend, Paul K.",
    title = "{On-shell versus Off-shell Equivalence in 3D Gravity}",
    eprint = "1812.09205",
    archivePrefix = "arXiv",
    primaryClass = "hep-th",
    doi = "10.1088/1361-6382/ab10e7",
    journal = "Class. Quant. Grav.",
    volume = "36",
    number = "9",
    pages = "095013",
    year = "2019"
}

@article{Ozkan:2018cxj,
    author = {\"Ozkan, Mehmet and Pang, Yi and Townsend, Paul K.},
    title = "{Exotic Massive 3D Gravity}",
    eprint = "1806.04179",
    archivePrefix = "arXiv",
    primaryClass = "hep-th",
    doi = "10.1007/JHEP08(2018)035",
    journal = "JHEP",
    volume = "08",
    pages = "035",
    year = "2018"
}

@article{Afshar:2019npk,
    author = "Afshar, Hamid Reza and Deger, Nihat Sadik",
    title = "{Exotic massive 3D gravities from truncation}",
    eprint = "1909.06305",
    archivePrefix = "arXiv",
    primaryClass = "hep-th",
    doi = "10.1007/JHEP11(2019)145",
    journal = "JHEP",
    volume = "11",
    pages = "145",
    year = "2019"
}

@article{Arvanitakis:2015oga,
	author = "Arvanitakis, Alex S. and Sevrin, Alexander and Townsend, Paul K.",
	title = "{Yang-Mills as massive Chern-Simons theory: a third way to three-dimensional gauge theories}",
	eprint = "1501.07548",
	archivePrefix = "arXiv",
	primaryClass = "hep-th",
	reportNumber = "DAMTP-2015-8",
	doi = "10.1103/PhysRevLett.114.181603",
	journal = "Phys. Rev. Lett.",
	volume = "114",
	number = "18",
	pages = "181603",
	year = "2015"
}

@article{Buchbinder:2024tui,
    author = "Buchbinder, I. L. and Kuzenko, S. M.",
    title = "{Quantum equivalence of the Freedman-Townsend model and the principal chiral $\sigma$-model}",
    eprint = "2405.16782",
    archivePrefix = "arXiv",
    primaryClass = "hep-th",
    month = "5",
    year = "2024"
}

@article{Tong:2022gpg,
    author = "Tong, David",
    title = "{A gauge theory for shallow water}",
    eprint = "2209.10574",
    archivePrefix = "arXiv",
    primaryClass = "hep-th",
    doi = "10.21468/SciPostPhys.14.5.102",
    journal = "SciPost Phys.",
    volume = "14",
    number = "5",
    pages = "102",
    year = "2023"
}

@article{Rosly:1996vr,
    author = "Rosly, A. A. and Selivanov, K. G.",
    title = "{On amplitudes in selfdual sector of Yang-Mills theory}",
    eprint = "hep-th/9611101",
    archivePrefix = "arXiv",
    reportNumber = "ITEP-TH-96-50",
    doi = "10.1016/S0370-2693(97)00268-2",
    journal = "Phys. Lett. B",
    volume = "399",
    pages = "135--140",
    year = "1997"
}

@article{Rosly:1997ap,
    author = "Rosly, A. A. and Selivanov, K. G.",
    title = "{Gravitational SD perturbiner}",
    eprint = "hep-th/9710196",
    archivePrefix = "arXiv",
    reportNumber = "ITEP-TH-56-97, IFUM-590-FT",
    year = "1997"
}

@article{Mafra:2015gia,
    author = "Mafra, Carlos R. and Schlotterer, Oliver",
    title = "{Solution to the nonlinear field equations of ten dimensional supersymmetric Yang-Mills theory}",
    eprint = "1501.05562",
    archivePrefix = "arXiv",
    primaryClass = "hep-th",
    reportNumber = "AEI-2015-005, DAMTP-2015-5",
    doi = "10.1103/PhysRevD.92.066001",
    journal = "Phys. Rev. D",
    volume = "92",
    number = "6",
    pages = "066001",
    year = "2015"
}

@article{Lee:2015upy,
    author = "Lee, Seungjin and Mafra, Carlos R. and Schlotterer, Oliver",
    title = "{Non-linear gauge transformations in $D=10$ SYM theory and the BCJ duality}",
    eprint = "1510.08843",
    archivePrefix = "arXiv",
    primaryClass = "hep-th",
    reportNumber = "DAMTP-2015-68",
    doi = "10.1007/JHEP03(2016)090",
    journal = "JHEP",
    volume = "03",
    pages = "090",
    year = "2016"
}

@article{Mafra:2015vca,
    author = "Mafra, Carlos R. and Schlotterer, Oliver",
    title = "{Berends-Giele recursions and the BCJ duality in superspace and components}",
    eprint = "1510.08846",
    archivePrefix = "arXiv",
    primaryClass = "hep-th",
    reportNumber = "DAMTP-2015-69",
    doi = "10.1007/JHEP03(2016)097",
    journal = "JHEP",
    volume = "03",
    pages = "097",
    year = "2016"
}

@article{Lee:2022aiu,
    author = "Lee, Kanghoon",
    title = "{Quantum off-shell recursion relation}",
    eprint = "2202.08133",
    archivePrefix = "arXiv",
    primaryClass = "hep-th",
    doi = "10.1007/JHEP05(2022)051",
    journal = "JHEP",
    volume = "05",
    pages = "051",
    year = "2022"
}

@article{Gomez:2022dzk,
    author = "Gomez, Humberto and Lipinski Jusinskas, Renann and Lopez-Arcos, Cristhiam and Quintero Velez, Alexander",
    title = "{One-Loop Off-Shell Amplitudes from Classical Equations of Motion}",
    eprint = "2208.02831",
    archivePrefix = "arXiv",
    primaryClass = "hep-th",
    doi = "10.1103/PhysRevLett.130.081601",
    journal = "Phys. Rev. Lett.",
    volume = "130",
    number = "8",
    pages = "081601",
    year = "2023"
}

@article{Gomez:2024xec,
    author = "Gomez, Humberto and Lipinski Jusinskas, Renann and Lopez-Arcos, Cristhiam and Quintero Velez, Alexander",
    title = "{One-loop $N$-point correlators in pure gravity}",
    eprint = "2411.07939",
    archivePrefix = "arXiv",
    primaryClass = "hep-th",
    month = "11",
    year = "2024"
}

@article{Gaiotto:2014kfa,
    author = "Gaiotto, Davide and Kapustin, Anton and Seiberg, Nathan and Willett, Brian",
    title = "{Generalized Global Symmetries}",
    eprint = "1412.5148",
    archivePrefix = "arXiv",
    primaryClass = "hep-th",
    doi = "10.1007/JHEP02(2015)172",
    journal = "JHEP",
    volume = "02",
    pages = "172",
    year = "2015"
}

@article{Lake:2018dqm,
    author = "Lake, Ethan",
    title = "{Higher-form symmetries and spontaneous symmetry breaking}",
    eprint = "1802.07747",
    archivePrefix = "arXiv",
    primaryClass = "hep-th",
    month = "2",
    year = "2018"
}

@article{Armstrong:2022mfr,
    author = "Armstrong, Connor and Gomez, Humberto and Lipinski Jusinskas, Renann and Lipstein, Arthur and Mei, Jiajie",
    title = "{New recursion relations for tree-level correlators in anti\textendash{}de Sitter spacetime}",
    eprint = "2209.02709",
    archivePrefix = "arXiv",
    primaryClass = "hep-th",
    doi = "10.1103/PhysRevD.106.L121701",
    journal = "Phys. Rev. D",
    volume = "106",
    number = "12",
    pages = "L121701",
    year = "2022"
}

@article{Argyres:1995jj,
    author = "Argyres, Philip C. and Douglas, Michael R.",
    title = "{New phenomena in SU(3) supersymmetric gauge theory}",
    eprint = "hep-th/9505062",
    archivePrefix = "arXiv",
    reportNumber = "IASSNS-HEP-95-31, RU-95-28",
    doi = "10.1016/0550-3213(95)00281-V",
    journal = "Nucl. Phys. B",
    volume = "448",
    pages = "93--126",
    year = "1995"
}

@article{Minahan:1996fg,
    author = "Minahan, Joseph A. and Nemeschansky, Dennis",
    title = "{An N=2 superconformal fixed point with E(6) global symmetry}",
    eprint = "hep-th/9608047",
    archivePrefix = "arXiv",
    reportNumber = "USC-96-18",
    doi = "10.1016/S0550-3213(96)00552-4",
    journal = "Nucl. Phys. B",
    volume = "482",
    pages = "142--152",
    year = "1996"
}

@article{Xie:2012hs,
    author = "Xie, Dan",
    title = "{General Argyres-Douglas Theory}",
    eprint = "1204.2270",
    archivePrefix = "arXiv",
    primaryClass = "hep-th",
    doi = "10.1007/JHEP01(2013)100",
    journal = "JHEP",
    volume = "01",
    pages = "100",
    year = "2013"
}

@article{Beem:2014zpa,
    author = "Beem, Christopher and Lemos, Madalena and Liendo, Pedro and Rastelli, Leonardo and van Rees, Balt C.",
    title = "{The $ \mathcal{N}=2 $ superconformal bootstrap}",
    eprint = "1412.7541",
    archivePrefix = "arXiv",
    primaryClass = "hep-th",
    reportNumber = "HU-EP-14-61, YITP-SB-14-54, CERN-PH-TH-2014-269, HU-EP-14/61",
    doi = "10.1007/JHEP03(2016)183",
    journal = "JHEP",
    volume = "03",
    pages = "183",
    year = "2016"
}

@article{Wang:2015mra,
    author = "Wang, Yifan and Xie, Dan",
    title = "{Classification of Argyres-Douglas theories from M5 branes}",
    eprint = "1509.00847",
    archivePrefix = "arXiv",
    primaryClass = "hep-th",
    reportNumber = "MIT-CTP-4711",
    doi = "10.1103/PhysRevD.94.065012",
    journal = "Phys. Rev. D",
    volume = "94",
    number = "6",
    pages = "065012",
    year = "2016"
}

@article{Garcia-Etxebarria:2015wns,
    author = "Garc\'\i{}a-Etxebarria, I\~naki and Regalado, Diego",
    title = "{$ \mathcal{N}=3 $ four dimensional field theories}",
    eprint = "1512.06434",
    archivePrefix = "arXiv",
    primaryClass = "hep-th",
    reportNumber = "MPP-2015-307",
    doi = "10.1007/JHEP03(2016)083",
    journal = "JHEP",
    volume = "03",
    pages = "083",
    year = "2016"
}

@article{Aharony:2015oyb,
    author = "Aharony, Ofer and Evtikhiev, Mikhail",
    title = "{On four dimensional N = 3 superconformal theories}",
    eprint = "1512.03524",
    archivePrefix = "arXiv",
    primaryClass = "hep-th",
    reportNumber = "WIS-11-15-NOV-DPPA",
    doi = "10.1007/JHEP04(2016)040",
    journal = "JHEP",
    volume = "04",
    pages = "040",
    year = "2016"
}

@article{Lemos:2016xke,
    author = "Lemos, Madalena and Liendo, Pedro and Meneghelli, Carlo and Mitev, Vladimir",
    title = "{Bootstrapping $\mathcal{N}=3$ superconformal theories}",
    eprint = "1612.01536",
    archivePrefix = "arXiv",
    primaryClass = "hep-th",
    reportNumber = "DESY-16-237, MITP-16-132, DESY 16-237",
    doi = "10.1007/JHEP04(2017)032",
    journal = "JHEP",
    volume = "04",
    pages = "032",
    year = "2017"
}

@article{Aharony:2016kai,
    author = "Aharony, Ofer and Tachikawa, Yuji and Gomi, Kiyonori",
    title = "{S-folds and 4d N=3 superconformal field theories}",
    eprint = "1602.08638",
    archivePrefix = "arXiv",
    primaryClass = "hep-th",
    reportNumber = "WIS-02-16-FEB-DPPA, IPMU-16-0022, UT-16-9, WIS/02/16-FEB-DPPA, IPMU-16-0022, UT-16-9",
    doi = "10.1007/JHEP06(2016)044",
    journal = "JHEP",
    volume = "06",
    pages = "044",
    year = "2016"
}

@book{Weinberg:2000cr,
    author = "Weinberg, Steven",
    title = "{The quantum theory of fields. Vol. 3: Supersymmetry}",
    isbn = "978-0-521-67055-5, 978-1-139-63263-8, 978-0-521-67055-5",
    publisher = "Cambridge University Press",
    month = "6",
    year = "2013"
}

@inproceedings{Witten:1995zh,
    author = "Witten, Edward",
    title = "{Some comments on string dynamics}",
    booktitle = "{STRINGS 95: Future Perspectives in String Theory}",
    eprint = "hep-th/9507121",
    archivePrefix = "arXiv",
    reportNumber = "IASSNS-HEP-95-63",
    pages = "501--523",
    month = "7",
    year = "1995"
}

@article{witten:2008,
      title={Conformal Field Theory In Four And Six Dimensions}, 
      author={Edward Witten},
      year={2008},
      eprint={0712.0157},
      archivePrefix={arXiv},
      primaryClass={math.RT},
      url={https://arxiv.org/abs/0712.0157}, 
}

@article{Witten:2009at,
    author = "Witten, Edward",
    title = "{Geometric Langlands From Six Dimensions}",
    eprint = "0905.2720",
    archivePrefix = "arXiv",
    primaryClass = "hep-th",
    month = "5",
    year = "2009"
}

@article{Beem:2015aoa,
    author = "Beem, Christopher and Lemos, Madalena and Rastelli, Leonardo and van Rees, Balt C.",
    title = "{The (2, 0) superconformal bootstrap}",
    eprint = "1507.05637",
    archivePrefix = "arXiv",
    primaryClass = "hep-th",
    reportNumber = "CERN-PH-TH-2015-165, YITP-SB-15-25",
    doi = "10.1103/PhysRevD.93.025016",
    journal = "Phys. Rev. D",
    volume = "93",
    number = "2",
    pages = "025016",
    year = "2016"
}

@article{Heckman:2018jxk,
    author = "Heckman, Jonathan J. and Rudelius, Tom",
    title = "{Top Down Approach to 6D SCFTs}",
    eprint = "1805.06467",
    archivePrefix = "arXiv",
    primaryClass = "hep-th",
    doi = "10.1088/1751-8121/aafc81",
    journal = "J. Phys. A",
    volume = "52",
    number = "9",
    pages = "093001",
    year = "2019"
}

@article{Freedman:1980us,
    author = "Freedman, Daniel Z. and Townsend, P. K.",
    title = "{Antisymmetric Tensor Gauge Theories and Nonlinear Sigma Models}",
    reportNumber = "ITP-SB-80-25",
    doi = "10.1016/0550-3213(81)90392-8",
    journal = "Nucl. Phys. B",
    volume = "177",
    pages = "282--296",
    year = "1981"
}

@article{Wyld:1961gqg,
    author = "Wyld, H. W",
    title = "{Formulation of the theory of turbulence in an incompressible fluid}",
    doi = "10.1016/0003-4916(61)90056-2",
    journal = "Annals Phys.",
    volume = "14",
    pages = "143--165",
    year = "1961"
}

@article{Cheung:2020djz,
    author = "Cheung, Clifford and Mangan, James",
    title = "{Scattering Amplitudes and the Navier-Stokes Equation}",
    eprint = "2010.15970",
    archivePrefix = "arXiv",
    primaryClass = "hep-th",
    reportNumber = "CALT-TH 2020-044",
    month = "10",
    year = "2020"
}

@article{Escudero:2022zdz,
    author = "Escudero, Valentina Guarin and Lopez-Arcos, Cristhiam and Quintero Velez, Alexander",
    title = "{Homotopy double copy and the Kawai\textendash{}Lewellen\textendash{}Tye relations for the non-abelian and tensor Navier\textendash{}Stokes equations}",
    eprint = "2201.06047",
    archivePrefix = "arXiv",
    primaryClass = "math-ph",
    doi = "10.1063/5.0119508",
    journal = "J. Math. Phys.",
    volume = "64",
    number = "3",
    pages = "032304",
    year = "2023"
}

@article{Kleiss:1988ne,
    author = "Kleiss, Ronald and Kuijf, Hans",
    title = "{Multi - Gluon Cross-sections and Five Jet Production at Hadron Colliders}",
    reportNumber = "Print-88-0425 (LEIDEN)",
    doi = "10.1016/0550-3213(89)90574-9",
    journal = "Nucl. Phys. B",
    volume = "312",
    pages = "616--644",
    year = "1989"
}

@article{Travaglini:2022uwo,
    author = "Travaglini, Gabriele and others",
    title = "{The SAGEX review on scattering amplitudes}",
    eprint = "2203.13011",
    archivePrefix = "arXiv",
    primaryClass = "hep-th",
    reportNumber = "SAGEX-22-01",
    doi = "10.1088/1751-8121/ac8380",
    journal = "J. Phys. A",
    volume = "55",
    number = "44",
    pages = "443001",
    year = "2022"
}

@article{Brandhuber:2022qbk,
    author = "Brandhuber, Andreas and Plefka, Jan and Travaglini, Gabriele",
    title = "{The SAGEX Review on Scattering Amplitudes Chapter 1: Modern Fundamentals of Amplitudes}",
    eprint = "2203.13012",
    archivePrefix = "arXiv",
    primaryClass = "hep-th",
    reportNumber = "SAGEX-22-02, HU-EP-22/06, QMUL-PH-22-01",
    doi = "10.1088/1751-8121/ac8254",
    journal = "J. Phys. A",
    volume = "55",
    number = "44",
    pages = "443002",
    year = "2022"
}

@article{Bern:2022wqg,
    author = "Bern, Zvi and Carrasco, John Joseph and Chiodaroli, Marco and Johansson, Henrik and Roiban, Radu",
    title = "{The SAGEX review on scattering amplitudes Chapter 2: An invitation to color-kinematics duality and the double copy}",
    eprint = "2203.13013",
    archivePrefix = "arXiv",
    primaryClass = "hep-th",
    reportNumber = "SAGEX-22-03, UUITP-18/22",
    doi = "10.1088/1751-8121/ac93cf",
    journal = "J. Phys. A",
    volume = "55",
    number = "44",
    pages = "443003",
    year = "2022"
}

@article{Bern:1994zx,
    author = "Bern, Zvi and Dixon, Lance J. and Dunbar, David C. and Kosower, David A.",
    title = "{One loop n point gauge theory amplitudes, unitarity and collinear limits}",
    eprint = "hep-ph/9403226",
    archivePrefix = "arXiv",
    reportNumber = "SLAC-PUB-6415, SACLAY-SPH-T-94-20, UCLA-TEP-94-4, SWAT-94-17",
    doi = "10.1016/0550-3213(94)90179-1",
    journal = "Nucl. Phys. B",
    volume = "425",
    pages = "217--260",
    year = "1994"
}

@article{Bern:1994cg,
    author = "Bern, Zvi and Dixon, Lance J. and Dunbar, David C. and Kosower, David A.",
    title = "{Fusing gauge theory tree amplitudes into loop amplitudes}",
    eprint = "hep-ph/9409265",
    archivePrefix = "arXiv",
    reportNumber = "SLAC-PUB-6563, SACLAY-SPH-T-94-95, UCLA-TEP-94-29, SWAT-94-36",
    doi = "10.1016/0550-3213(94)00488-Z",
    journal = "Nucl. Phys. B",
    volume = "435",
    pages = "59--101",
    year = "1995"
}

@article{Bern:2011qt,
    author = "Bern, Zvi and Huang, Yu-tin",
    title = "{Basics of Generalized Unitarity}",
    eprint = "1103.1869",
    archivePrefix = "arXiv",
    primaryClass = "hep-th",
    reportNumber = "UCLA-11-TEP-103",
    doi = "10.1088/1751-8113/44/45/454003",
    journal = "J. Phys. A",
    volume = "44",
    pages = "454003",
    year = "2011"
}

@article{Gibbons:1995vg,
    author = "Gibbons, Gary W. and Green, Michael B. and Perry, Malcolm J.",
    title = "{Instantons and seven-branes in type IIB superstring theory}",
    eprint = "hep-th/9511080",
    archivePrefix = "arXiv",
    reportNumber = "DAMTP-R-95-56",
    doi = "10.1016/0370-2693(95)01565-5",
    journal = "Phys. Lett. B",
    volume = "370",
    pages = "37--44",
    year = "1996"
}

@article{Gubser:1996wt,
    author = "Gubser, Steven S. and Hashimoto, A. and Klebanov, Igor R. and Maldacena, Juan Martin",
    title = "{Gravitational lensing by p-branes}",
    eprint = "hep-th/9601057",
    archivePrefix = "arXiv",
    reportNumber = "PUPT-1586",
    doi = "10.1016/0550-3213(96)00182-4",
    journal = "Nucl. Phys. B",
    volume = "472",
    pages = "231--248",
    year = "1996"
}

\end{document}